\documentclass[twocolumn,superscriptaddress,showpacs,nofootinbib,preprintnumbers,secnumarabic,amssymb, nobibnotes, aps, prd]{revtex4-2}
\usepackage[utf8]{inputenc}
\usepackage{graphicx}
\usepackage{latexsym,amsmath,amssymb,amsthm,lmodern,float,url}
\usepackage{natbib}
\usepackage{color}
\usepackage{microtype}
\usepackage{import}
\usepackage{bbold}
\usepackage[plain]{fancyref}
\usepackage{varioref}
\usepackage{slashed}
\usepackage{multirow}
\usepackage{tikz}
\usepackage{scrextend}
\usepackage{braket}
\usetikzlibrary{shapes}
\usetikzlibrary{positioning}
\usepackage[normalem]{ulem}

\usepackage{qcircuit}

\usepackage[colorlinks=true,backref=false, linktocpage=true,
citecolor=blue,urlcolor=blue,linkcolor=blue,pdfpagemode=UseOutlines]{hyperref}
\hypersetup{%
  bookmarksnumbered=true,
  pdftitle = {},
  pdfsubject = {},
  pdfauthor = {},
  pdfkeywords = {}
}

\let\Re\undefined

\DeclareMathOperator{\Tr}{Tr}
\DeclareMathOperator{\Re}{Re}

\DeclareMathOperator{\tr}{Tr}
\DeclareMathOperator{\re}{Re}

\newcommand{\bi}{\mathbb{BI}}

\newcommand{\btt}{\mathbb{BT}}

\newcommand{\qit}{quicosotetrit }
\newcommand{\qits}{quicosotetrits }

\definecolor{blugrn}{RGB}{0,158,115}

\begin{document}
\preprint{FERMILAB-PUB-22-583-SQMS-T}
\title{Primitive Quantum Gates for an \texorpdfstring{$SU(2)$}{SU(2)} Discrete Subgroup: \texorpdfstring{$\mathbb{BT}$}{BT}}
\author{Erik J. Gustafson}
\email{egustafs@fnal.gov}
\affiliation{Fermi National Accelerator Laboratory, Batavia,  Illinois, 60510, USA}
\author{Henry Lamm}
\email{hlamm@fnal.gov}
\affiliation{Fermi National Accelerator Laboratory, Batavia,  Illinois, 60510, USA}
\author{Felicity Lovelace}
\email{fl16@uic.edu}
\affiliation{Department of Physics, University of Illinois at Chicago, Chicago, Illinois 60607, USA}
\author{Damian Musk}
\email{dmusk@ohs.stanford.edu}
\affiliation{Stanford University Online High School, Redwood City, CA 94063, USA}
\date{\today}

\begin{abstract}
We construct a primitive gate set for the digital quantum simulation of the binary tetrahedral ($\mathbb{BT}$) group on two quantum architectures. This nonabelian discrete group serves as a crude approximation to $SU(2)$ lattice gauge theory while requiring five qubits or one quicosotetrit per gauge link. The necessary basic primitives are the inversion gate, the group multiplication gate, the trace gate, and the $\mathbb{BT}$ Fourier transform over $\mathbb{BT}$. We experimentally benchmark the inversion and trace gates on \texttt{ibm\_nairobi}, with estimated fidelities between $14-55\%$, depending on the input state.

\end{abstract}

\maketitle

\section{Introduction}

Simulating the dynamics of lattice field theories offers a clear potential for quantum advantage~\cite{Feynman:1981tf,Lloyd1073,Jordan:2011ne,Jordan:2017lea,klco2021standard,Bauer:2022hpo}. Time evolution on quantum computers requires efficiently implementing the unitary operator $U(t)=e^{-iHt}$. Various approximations for $U(t)$ with different tradeoffs exist~\cite{Jordan:2011ne,Bender:2018rdp,haah2018quantum,Du:2020glq,PhysRevX.11.011020,PhysRevLett.123.070503,PhysRevLett.123.070503,berry2009black,PhysRevLett.114.090502,Low2019hamiltonian,Low2019hamiltonian,PhysRevLett.118.010501,cirstoiu2020variational,gibbs2021longtime,yao2020adaptive}, but all of them require implementing key group theoretic operations as quantum circuits in the case of lattice gauge theories~\cite{Lamm:2019bik}. This separation of the problem into group-dependent primitives~\cite{Alam:2021uuq} and algorithmic design~\cite{Carena:2022kpg} has proven fruitful in optimizing both.

For efficient digital simulations, many proposals exist on how the lattice gauge degrees of freedom can be truncated~\cite{Zohar:2012ay,Zohar:2012xf,Zohar:2013zla,Zohar:2014qma,Zohar:2015hwa,Zohar:2016iic,Klco:2019evd,Ciavarella:2021nmj,Bender:2018rdp,Liu:2020eoa,Hackett:2018cel,Alexandru:2019nsa,Yamamoto:2020eqi,Haase:2020kaj,Armon:2021uqr,PhysRevD.99.114507,Bazavov:2015kka,Zhang:2018ufj,Unmuth-Yockey:2018ugm,Unmuth-Yockey:2018xak,Kreshchuk:2020dla,Kreshchuk:2020aiq,Raychowdhury:2018osk,Raychowdhury:2019iki,Davoudi:2020yln,Wiese:2014rla,Luo:2019vmi,Brower:2020huh,Mathis:2020fuo,Singh:2019jog,Singh:2019uwd,Buser:2020uzs,Bhattacharya:2020gpm,Barata:2020jtq,Kreshchuk:2020kcz,Ji:2020kjk,Bauer:2021gek,Gustafson:2021qbt,Hartung:2022hoz,Grabowska:2022uos,Murairi:2022zdg}.  For some regulated theories, the desired theory may not even be the true continuum limit~\cite{Hasenfratz:2001iz,Caracciolo:2001jd,Hasenfratz:2000hd,PhysRevE.57.111,PhysRevE.94.022134,car_article,Singh:2019jog,Singh:2019uwd,Bhattacharya:2020gpm,Zhou:2021qpm,Caspar:2022llo}. Furthermore, the relative efficacy of schemes is dimension-dependent~\cite{Davoudi:2020yln,Zohar:2021nyc,Alam:2021uuq}.

One promising digitization method is the discrete subgroup approximation~\cite{Bender:2018rdp,Hackett:2018cel,Alexandru:2019nsa,Yamamoto:2020eqi,Ji:2020kjk,Haase:2020kaj,Carena:2021ltu,Armon:2021uqr,Gonzalez-Cuadra:2022hxt}. This method was explored early on in Euclidean lattice field theory to reduce resources. Replacing $U(1)$ by $\mathbb{Z}_N$ was considered in~\cite{Creutz:1979zg,Creutz:1982dn}. Extensions to the crystal-like subgroups of $SU(N)$ were made in Refs. ~\cite{Bhanot:1981xp,Petcher:1980cq,Bhanot:1981pj,Hackett:2018cel,Alexandru:2019nsa,Ji:2020kjk,Ji:2022qvr,Alexandru:2021jpm,Carena:2022hpz}, including with fermions~\cite{Weingarten:1980hx,Weingarten:1981jy}. Theoretical studies revealed that the discrete subgroup approximation corresponds to continuous groups broken by a Higgs mechanism~\cite{Kogut:1980qb,romers2007discrete,Fradkin:1978dv,Harlow:2018tng,Horn:1979fy}.  On the lattice, this causes the discrete subgroup to poorly approximate the continuous group below a \emph{freezeout} lattice spacing $a_f$ (or beyond a coupling $\beta_f$).

Lattice calculations are performed at fixed lattice spacing $a=a(\beta)$ which approaches zero as $\beta\rightarrow\infty$ for asymptotically free theories. Finite $a$ leads to discrepancies from the continuum results, but provided one simulates in the \emph{scaling regime} below $a_s(\beta_s)$, these errors should be polynomial in $a$.  Any approximation error from using discrete subgroups should be tolerable provided $a_s\gtrsim a_f$ or equivalently that $\beta_s\lesssim\beta_f$. For the $3+1d$ Wilson action, $\beta_f$ are known. In the case of $U(1)$ where $\beta_s= 1$, $\mathbb{Z}_{n>5}$ satisfies $\beta_f>\beta_s$. For nonabelian gauge groups, only a few crystal-like subgroups exist. $SU(2)$ has three: the binary tetrahedral $\mathbb{BT}$, the binary octahedral $\mathbb{BO}$, and the binary icosahedral $\bi$. 
The scaling regime for $SU(2)$ occurs around $\beta_s=2.2$.  Therefore, a value of $\beta_f=2.24(8)$ for $\mathbb{BT}$ is unlikely to prove useful with just the Kogut-Susskind Hamiltonian $H_{KS}$, although experience with $SU(3)$ suggests modified or improved Hamiltonians $H_{I}$ would prove sufficient~\cite{Alexandru:2019nsa,Ji:2020kjk,Ji:2022qvr,Alexandru:2021jpm}. The other two groups, $\mathbb{BO}$  and $\mathbb{BI}$, have values far into the scaling regime: $\beta_f=3.26(8)$ and $\beta_f=5.82(8)$, respectively~\cite{Alexandru:2019nsa}.

Substantial work has studied the quantum simulation of abelian theories, particularly in low dimensions. Despite this, one must remember that nonabelian gauge theories demonstrate many behaviors unseen in abelian ones; thus, results for $U(1)$ or $\mathbb{Z}_N$ may fail to represent the full complexity of lattice gauge theories. The group of interest in this paper, the 24-element $\mathbb{BT}$, is the smallest crystal-like subgroup of a nonabelian theory and requires 5 qubits per register. The dihedral groups, $D_N$, while not crystal-like, have previously been investigated for simulation on quantum computers~\cite{Bender:2018rdp,Lamm:2019bik,Alam:2021uuq,Fromm:2022vaj}.  Having $2N$ elements respectively, they require $\lceil\log_2(2N)\rceil$ qubits per register. Further studies have been undertaken to understand the $\mathbb{Q}_8$ subgroup of $SU(2)$~\cite{Gonzalez-Cuadra:2022hxt} which requires only 3 qubits.

\begin{figure}[!th]
\centering
    \includegraphics[width=0.9\linewidth]{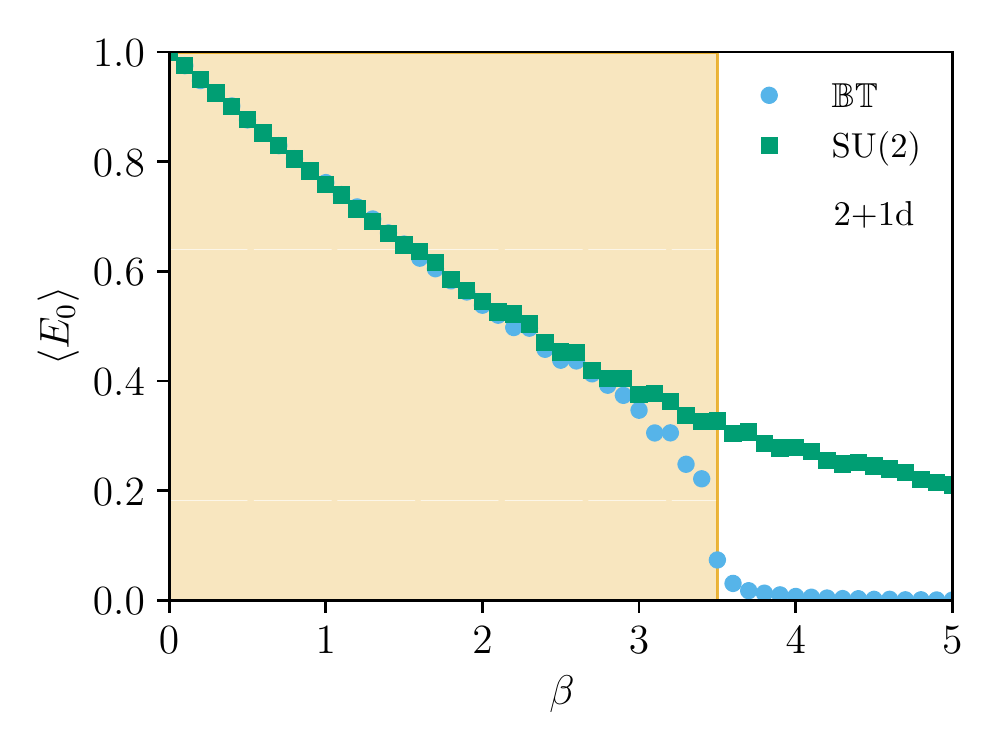}
    \includegraphics[width=0.9\linewidth]{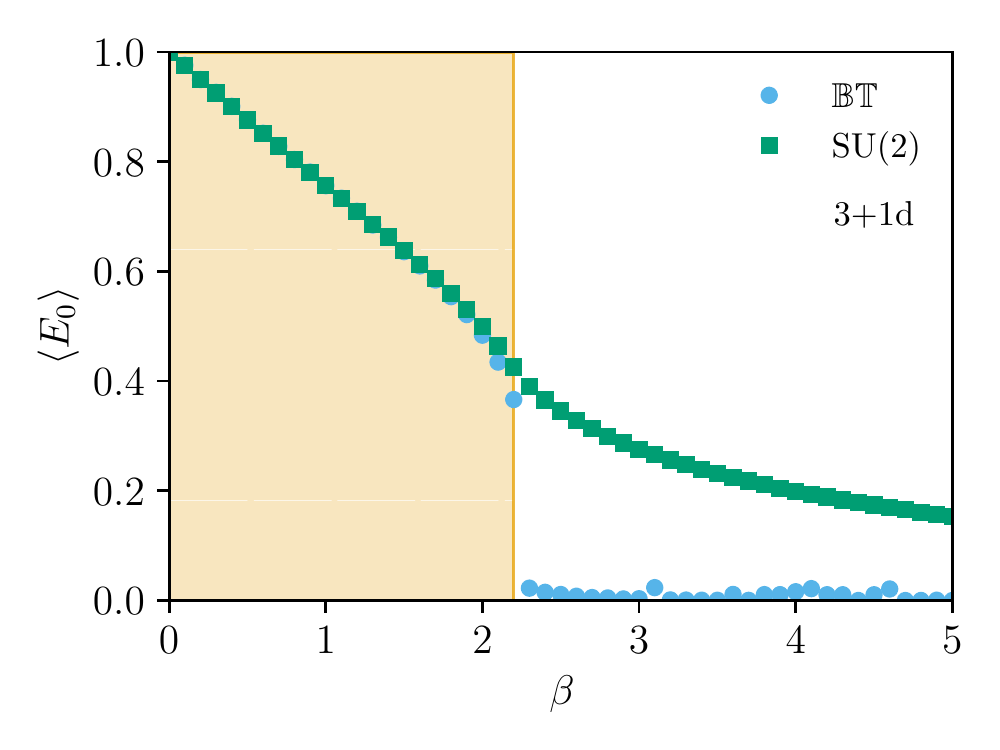}
    \caption{\label{fig:freezing}Euclidean calculations of lattice energy density $\langle E_0\rangle$ of $\btt$ as measured by the expectation value of the plaquette as a function of Wilson coupling $\beta$  on $4^d$ lattices for (top) $2+1d$ (bottom) $3+1d$. The shaded region indicates $\beta\leq \beta_s$.} 
\end{figure}

In the interest of studying quantum simulations on near-term devices, we should consider both $3+1d$ and $2+1d$ theories. Using classical lattice simulations, we have determined that in both spacetimes $\beta_f\approx\beta_s$ for the Wilson action (See Fig.~\ref{fig:freezing}). Thus quantum simulations with $\btt$ require an improved Hamiltonian~\cite{Carena:2022kpg} and will be the only one considered in this work. Since $a_f\propto e^{-\beta_f}$ within the scaling regime, only a small improvement in the Hamiltonian is needed. 

In this paper, we construct quantum circuits 
implementing the four primitive gates (inversion, multiplication, trace, and Fourier) required to simulate the $\btt$ theories. We will consider two possible quantum devices when constructing our gates.  The first is a qubit-based device.  The second device, motivated by the potential for bosonic quantum comptuers~\cite{Alam:2022crs}, is a $d=24$ qudit device where only one qudit is required per register. We refer to this d=24 state qudit as a $\qit$. Although time evolution on quantum processors is infeasible at present due to the low gate fidelities and coherence time, we benchmark the process fidelity of the inversion and trace gates for $\btt$ on the \texttt{ibm\_nairobi} QPU to evaluate the improvements needed for simulations on quantum processors.

This paper is organized as follows. In Sec.~\ref{sec:group}, the necessary group theoretic properties of $\btt$ are summarized and the digitization scheme is presented. A review of the basic qubit and qudit gates used in this work is found in Sec.~\ref{sec:qgates}. Sec.~\ref{sec:gates} summarized the four primitive gates required for implementing the group operations. This is followed by
quantum circuit constructions for these gates for $\btt$ 
gauge theories: 
the inversion gate in Sec.~\ref{sec:inverse}, the multiplication gate in Sec.~\ref{sec:multiplication}, the trace gate in Sec.~\ref{sec:trace}, and the Fourier transform gate in Sec.~\ref{sec:fourier}. Benchmark results for our $\btt$ inversion and trace gates are found in Sec.~\ref{sec:results}.  Using these gates, Sec.~\ref{sec:resources} presents a resource estimates for simulating $3+1d$ $SU(2)$.  We conclude and discuss future work in Sec.~\ref{sec:conclusions}.

\section{\label{sec:group}Properties of \texorpdfstring{$\btt$}{BT}}
The simulation of lattice gauge theories requires the definition of a register where one can store the state of a bosonic link variable which we call a $G-$register. In order to construct the $\btt-$register in term of integers, it is necessary to construct a mapping between the 24 elements of the group and the integers $[0,23]$. A clean way to obtain this is to write every element of $\mathbb{BT}$ as an ordered product of four generators\footnote{The minimal set of generators for $\btt$ is two, but we have been unable to find an ordered product with less than three. The choice of three generators is the same as the one with four generators where $(-1)^m\mathbf{i}^n\rightarrow \mathbf{i}^{2m + n}$. Nevertheless the qubit costs cannot go below the current formulation's value of $\lceil\log_2(24)\rceil=5$.} with exponents written in terms of the binary variables $m,n,o,p,q$:
\begin{equation}
\label{eq:pres}
    g = (-1)^m \mathbf{i}^n \mathbf{j}^o \mathbf{l}^{p + 2q},
\end{equation}
with 
\begin{equation}
    \mathbf{l} = -\frac{1}{2} (\mathbb{1}+ \mathbf{i} + \mathbf{j} + \mathbf{k})
\end{equation}
and $\mathbf{i}$, $\mathbf{j}$, $\mathbf{k}$ are the unit quaternions which in the 2d irreducible representation (irrep) correspond to Pauli matrices.
With the construction of Eq.~(\ref{eq:pres}), the $\btt-$register is given by a binary encoding of the qubits with the ordering $|qponm\rangle$.  This same mapping can be utilized for \qits.
For example, using $\eta=1+i$ one element in the real 2d irrep is
\begin{equation}
\frac{1}{2}\begin{pmatrix}
\eta&\eta^*\\-\eta&\eta^*
\end{pmatrix}=(-1)^1\mathbf{i}^1\mathbf{j}^1\mathbf{l}^{0+2\times1}\rightarrow|10111\rangle=|23\rangle
\end{equation}

The $\mathbf{i}$, $\mathbf{j}$, and $\mathbf{k}$ generators anti-commute with each other. Additional useful relations are:
\begin{equation}
\label{eq:genrel}
\begin{split}
&\mathbf{i}^2=\mathbf{j}^2=-\mathbb{1},~ \mathbf{l}^3=\mathbb{1}\\
&\mathbf{i} \mathbf{j} = \mathbf{k}, ~ \mathbf{j} \mathbf{k} = \mathbf{i}, ~ \mathbf{k} \mathbf{i} = \mathbf{j},\\
&\mathbf{l}\mathbf{i}=\mathbf{j}\mathbf{l},~\mathbf{l}\mathbf{j}=\mathbf{k}\mathbf{l},~\mathbf{l}\mathbf{k}=\mathbf{i}\mathbf{l},\\
&\mathbf{l}^2\mathbf{i}=\mathbf{k}\mathbf{l}^2,~\mathbf{l}^2\mathbf{j}=\mathbf{k}\mathbf{l}^2,~\mathbf{l}^2\mathbf{k}=\mathbf{j}\mathbf{l}^2.\\
\end{split}
\end{equation}

\begin{table}[b]
    \caption{Character Table of $\mathbb{BT}$ including an enumeration of the elements in the given class.}
    \label{tab:charbt}
    \centering
    \begin{tabular}{c||c|c|c|c|c|c|c}
Size & 1 & 1 & 6 & 4 & 4 & 4 & 4 \\
Order & 1 & 2 & 4 & 6 & 6 & 3 & 3 \\
\hline\hline               
$\rho_1$ & 1 & 1 & 1 & 1 & 1 & 1 & 1 \\
$\rho_2$ & 1 & 1 & 1 & $\omega$ & $\omega^2$ & $\omega^2$ & $\omega$ \\
$\rho_3$ & 1 & 1 & 1 & $\omega^2$ & $\omega$ & $\omega$ & $\omega^2$ \\
$\rho_4$ & 2 & $-2$ & 0 & 1 & 1 & $-1$ & $-1$ \\
$\rho_5$ & 2 & $-2$ & 0 & $\omega^2$ & $\omega$ & $-\omega^2$ & $-\omega$ \\
$\rho_6$ & 2 & $-2$ & 0 & $\omega^2$ & $\omega$ & $-\omega$ & $-\omega^2$ \\
$\rho_7$ & 3 & 3 & $-1$ & 0 & 0 & 0 & 0 \\
\hline               
$\ket{g}$ & $\ket{0}$ & $\ket{1}$ & $\ket{2}$,$\ket{3}$ & $\ket{9}$,$\ket{10}$ & $\ket{17}$,$\ket{19}$ & $\ket{8}$,$\ket{11}$ & $\ket{16}$,$\ket{18}$ \\
\phantom{x} & \phantom{x} & \phantom{x} & $\ket{4}$,$\ket{5}$ & $\ket{12}$,$\ket{14}$ & $\ket{21}$,$\ket{23}$ & $\ket{13}$,$\ket{15}$ & $\ket{20}$,$\ket{22}$ \\
\phantom{x} & \phantom{x} & \phantom{x} & $\ket{6}$,$\ket{7}$ & \phantom{x} & \phantom{x} & \phantom{x} & \phantom{x} \\
    \end{tabular}
\end{table}

The character table (Table~\ref{tab:charbt}) lists important group properties; the different irreps can be identified by the value of their character acting on each element. An irrep's dimension is the value of the character of $\mathbb{1}$.  There are three $1d$ irreps, three $2d$ irreps (one real and two complex), and one $3d$ irrep.  To derive the Fourier transform, it is necessary to know a matrix presentation of each irrep. Based on our qubit mapping, given a presentation of $-1,\mathbf{i},\mathbf{j},$ and $\mathbf{l}$ we can construct any element of the group from Eq.~(\ref{eq:pres}). With the $3$rd root of unity $\omega=e^{2\pi i/3}$, the 1d irreps are given:
\begin{equation}
\label{eq:reps1}
    \rho_1: -1=\mathbf{i}=\mathbf{j}=\mathbf{l}=1
    \end{equation}
    \begin{equation}
    \rho_2: -1=\mathbf{i}=\mathbf{j}=1,\; \mathbf{l}=\omega^2
        \end{equation}
\begin{equation}
    \rho_3: -1=\mathbf{i}=\mathbf{j}=1,\; \mathbf{l}=\omega
    \end{equation}
Now for the 2d irreps, we can use for all three irreps the same definitions:
\begin{equation}
\label{eq:reps2}
\begin{split}
\rho_{4,5,6}:& -1=\text{diag}(-1,-1),\; \mathbf{i}=\text{diag}(i,-i),\\\;&\mathbf{j}=\begin{pmatrix}0&-1 \\1&0\end{pmatrix},\;
    \mathbf{l}=-\frac{1}{2}\begin{pmatrix}\eta&-\eta\\\eta^*&\eta^*\end{pmatrix}
    \end{split}
\end{equation}
then we can construct the three 2d irreps by taking:
\begin{equation}
    \rho_4(g) = (-1)^m \mathbf{i}^n \mathbf{j}^o \mathbf{l}^{p + 2q}
    \end{equation}
    \begin{equation}
    \rho_5(g) = (-1)^m \mathbf{i}^n \mathbf{j}^o (\omega^2 \mathbf{l})^{p + 2q}    
    \end{equation}
\begin{equation}
    \rho_6(g) = (-1)^m \mathbf{i}^n \mathbf{j}^o (\omega \mathbf{l})^{p + 2q}
\end{equation}

For the 3d irrep, we have
\begin{equation}
\label{eq:reps4}
    \begin{split}
    \rho_7&: -1=\text{diag}(1,1,1),\quad \mathbf{i}=\text{diag}(-1,1,-1),\\&\quad \mathbf{j}=\text{diag}(1,-1,-1),\quad
    \mathbf{l}=\begin{pmatrix}0&1&0\\0&0&1\\1&0&0\end{pmatrix}
\end{split}
\end{equation}

\section{Qubit and Qudit Gates}
\label{sec:qgates}

In order to implement the group primitive gates on qubit and qudit hardware we need a set of quantum gate operations. We begin by first enumerating the qubit gates, followed by a discussion of the qudit gates.

The first basic qubit gates we need are the Pauli gates $p=X,Y,Z$. These can be extended to arbitrary rotations about their respective axes $R_p(\theta)=e^{i\theta p/2}$.  When decomposing onto fault-tolerant devices, the $T=\text{diag}(1,e^{i\pi/4})$ gate becomes relevant.

The first multiqubit operation we need is the SWAP operation, which swaps two qubits:
        \begin{equation*}
            \text{SWAP} ~\ket{a}\otimes\ket{b} = \ket{b}\otimes\ket{a}.
        \end{equation*}
The controlled not (CNOT) gate applies the $X$ operation on a target qubit if the control qubit is in the state $\ket{1}$:
        \begin{equation*}
            \text{CNOT} \ket{a}\otimes\ket{b} = \ket{a}\otimes\ket{b\oplus a},
        \end{equation*}
        where $\oplus$ indicates addition modulus 2.
We also need the following multiqubit gates: C$^n$NOT -- of which C$^2$NOT is called the Toffoli gate -- and CSWAP (Fredkin) gates. The C$^n$NOT gate is the further extension to the case of where the $n$ control qubits must be in the $\ket{1}^{\otimes n}$ state. For example, the Toffoli in terms of modular arithmetic is
        \begin{equation*}
            \text{C}^2\text{NOT} \ket{a}\otimes\ket{b}\otimes\ket{c}=\ket{a}\otimes\ket{b}\otimes\ket{c \oplus ab}.
        \end{equation*}
The CSWAP gate swaps two qubit states if the control is in the $\ket{1}$ state:
        \begin{equation*}
        \begin{split}
            \text{CSWAP}\ket{a}\otimes\ket{b}\otimes\ket{c}=&\ket{a}\otimes\ket{b(1\oplus a)\oplus ac}\\&\otimes\ket{c(1\oplus a)\oplus a b}.\\
            \end{split}
        \end{equation*}

\begin{figure}[h]
 \includegraphics[width=0.62\linewidth]{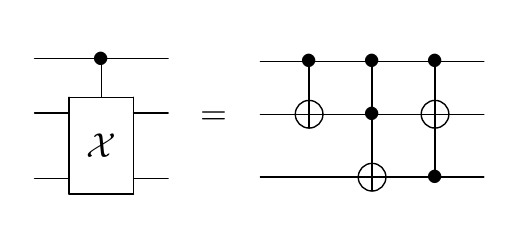}
 \caption{The controlled permutation gate, C$\chi$.}
    \label{fig:dcpgate}
\end{figure}

One final qubit operation, the controlled permutation gate $C\chi$, will prove useful to define for conciseness later.  Fig.~\ref{fig:dcpgate} constructs it in terms of C$^n$NOT gates.

We also need a set of gates for implementation on \qit device. In our case, there are not specialized \qit gates but a general set of qudit ones to consider. The single qudit gates we need are: Givens rotations, the selective number of arbitrary photon (SNAP)~ \cite{https://doi.org/10.48550/arxiv.2004.14256,Heeres_2015}, displacement~\cite{Krastanov_2015}, and photon blockade gates~\cite{chakram2022multimode,2019PhRvA.100f3817H}.

Givens rotations $R_{p}^{(a,b)}(\theta)$ are generalizations of $R_p(\theta)$ to qudits where rotations occur in the subspace of states $\ket{a}$ and $\ket{b}$ while leaving all the other states untouched. We also use the notation $p^{(a,b)}$ to indicate special Givens rotations that correspond to the generalized Pauli gates (e.g. $X^{(3,4)}$).  These gates are useful in designing algorithms for simulating with qudits, but are difficult to natively implement.  Therefore real simulations will likely require their decomposition.

One native gate set for single qudits in cavity QED devices is SNAP and displacement gates 
\cite{Heeres_2015, Krastanov_2015, https://doi.org/10.48550/arxiv.2004.14256}. SNAP gates can arbitrarily phase the qudit states:
\begin{equation}
    \mathcal{S}(\vec{\theta}) = \sum_{a=0}^{N-1} \ket{a}\bra{a} e^{i \theta_a}
\end{equation}
where the sum is over computational basis states $a=[0,N-1]$ of an $N-$state qudit and $\vec{\theta}=\{\theta_0,\theta_1,\hdots,\theta_{N-1}\}$ are a set of tunable parameters analogous to $\theta$ in $R_Z(\theta)$.

The displacement gate coherently changes the cavity's photon number. In terms of Fock-operators $\hat{a},\hat{a}^\dagger$, it is
\begin{equation}
    \mathcal{D}(\alpha) = e^{ \alpha\hat{a}^{\dagger}-\alpha^* \hat{a}}.
\end{equation}

The photon blockade operation acts as an $R_X(\theta)$ or $R_Y(\theta)$ rotation between two Fock states in a cavity by driving the system at an off-resonant frequency while shifting the desired modes to said frequency~\cite{chakram2022multimode, 2019PhRvA.100f3817H}.

In order to obtain a universal set of gates, we require an entangling gate. One proposal is the controlled SNAP gate which phases a target qudit based on if a second qudit is in a state $\ket{\alpha}$~\cite{PhysRevLett.127.107701}:
\begin{equation}
       \mathcal{S}_c(\vec{\theta}, \alpha) = \sum_{a = 0, a\neq \alpha}^{N-1} |a\rangle\langle a| \otimes \mathbb{1} + |\alpha\rangle \langle \alpha| \otimes \sum_{b=0}^{N-1}|b\rangle\langle b|e^{i \theta_b}.
\end{equation}

\section{\label{sec:gates}Overview of Primitive Gates}

For general gauge groups, it is possible to define any quantum circuit with sets of primitive gates.  Using this formulation confers two benefits: first, it is possible to design algorithms in a theory- and hardware-agnostic way; second, the circuit optimization is split into smaller, more manageable pieces. This construction begins with defining for $G$ a $G$-register by identifying each group element with a computational basis state $\ket{g}$, where  $g\in G$. One choice of primitive gates is: inversion $\mathfrak U_{-1}$, multiplication $\mathfrak U_{\times}$, trace $\mathfrak U_{\rm Tr}$, and Fourier transform $\mathfrak U_{F}$~\cite{Lamm:2019bik}.

The inversion gate, $\mathfrak U_{-1}$, is a single register gate which takes a group element to its inverse:
\begin{equation}
\mathfrak U_{-1} \left|g\right> = \left|g^{-1}\right>\text.
\end{equation}

The group multiplication gate acts on two $G-$registers. It takes the target $G-$register and changes the state to the left product with the control $G-$register:
\begin{equation}
    \mathfrak U_{\times} \ket{g}\ket{h} = \ket{g} \ket{gh}.
\end{equation}
Left multiplication is sufficient for a minimal set as right multiplication can be implemented use two applications of $\mathfrak U_{-1}$ and $\mathfrak U_{\times}$, albeit optimal algorithms may take advantage of an explicit construction~\cite{Carena:2022kpg}.

The trace of products of group elements appears in lattice Hamiltonians.  We can implement these terms by combining $\mathfrak U _{\times}$ with a single-register trace gate:
\begin{equation}
\mathfrak U_{\Tr}(\theta) \left|g\right> = e^{i \theta \Re\Tr g} \left|g\right>.
\label{eqn:trace-gate}
\end{equation}

\begin{figure}
    \centering
    \includegraphics[width=0.8\linewidth]{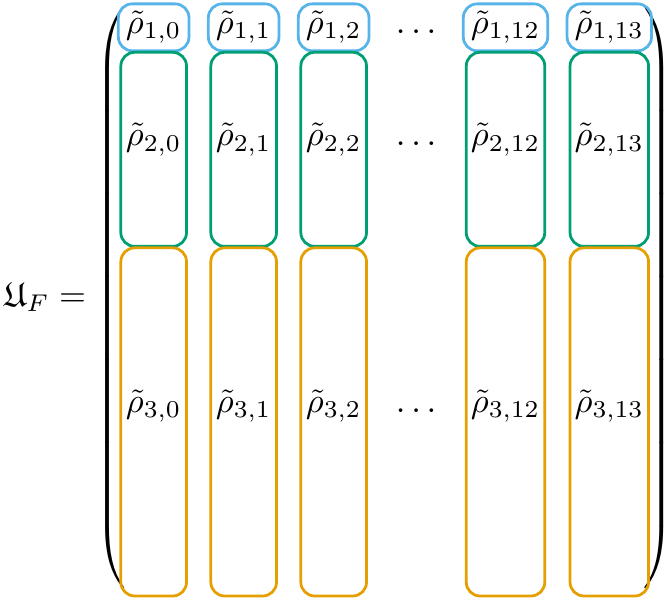}
    \caption{Example $\mathfrak U_{F}$ from Eq.~(\ref{eqn:Fourier-group}) using column vectors  $\tilde\rho_{i,j}=\sqrt{d_\rho/|G|}\rho_{i,j}$ where $\rho_{i,j}=\rho_i(g_j)$. This example has three irreps with $d_\rho=1,2,3$.  $\mathfrak U_{F}$ is square since $\sum_\rho d_{\rho}^2=|G|$}
    \label{fig:qft_cartoon}
\end{figure}

The final gate required is the group Fourier transform $\mathfrak U_F$. The Fourier transform of a finite $G$ is defined as
\begin{eqnarray}
\hat{f}(\rho) = \sqrt{\frac{d_{\rho}}{|G|}} \sum_{g \in G} f(g) \rho(g),
\label{eqn:Fourier-group}
\end{eqnarray}
where $\vert G \vert$ is the size of the group, $d_{\rho}$ is the dimensionality of the representation $\rho$, and $f$ is a function over $G$.
The inverse transform is given by
\begin{eqnarray}
\label{eq:dft}
f(g) = \frac{1}{\sqrt{|G|}} \sum_{\rho \in \hat{G}} \sqrt{d_{\rho}} \tr{(\hat{f}(\rho) \rho(g^{-1}))},
\end{eqnarray}
where the dual $\hat{G}$ is the set of all irreducible representations (irrep) of $G$. The gate that performs this acts on a single $G$-register with some amplitudes $f(g)$ which rotate it into the Fourier basis:
\begin{equation}
\label{eq:uft}
\mathfrak U_F \sum_{g \in G} f(g)\left|g \right>
=
\sum_{\rho \in \hat G} \hat f(\rho)_{ij} \left|\rho,i,j\right>.
\end{equation}
The second sum is taken over $\rho$, the irreducible representations of $G$; $\hat f$ denotes the Fourier transform of $f$. After application of the gate, the register is denoted as a $\hat G$-register to indicate the change of basis. A schematic example of this gate is show in Fig.~\ref{fig:qft_cartoon}

\section{\label{sec:inverse}Inversion Gate}
\begin{figure}
 \includegraphics[width=0.78\linewidth]{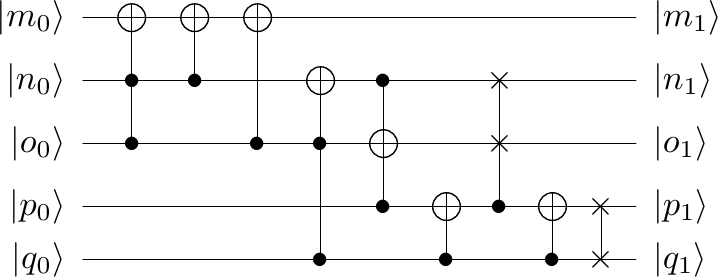}
    \caption{A qubit implementation of $\mathfrak U_{-1}$ which uses Toffoli and CSWAP gates.  These multiqubit entangling gates can be decomposed into one- and two-qubit gates as discussed in the literature.}
    \label{fig:inversiongate}
\end{figure}

Consider a $\btt-$register storing the group element given by $g=(-1)^{m_0}\mathbf{i}^{n_0}\mathbf{j}^{o_0}\mathbf{l}^{p_0 + 2q_0}$. The effect of the inversion gate on this register is to transform it to
\begin{equation}
    |g\rangle=|q_0p_0o_0n_0m_0\rangle\rightarrow |g^{-1}\rangle=|q_1p_1o_1n_1m_1\rangle.
\end{equation} 
Using Eq.~(\ref{eq:pres}) and $(AB)^{-1}=B^{-1}A^{-1}$, the inverse of $g$ is 
\begin{equation*}
    g^{-1} = (-1)^{m_0 + n_0 + o_0} \mathbf{l}^{3 - p_0 - 2q_0} \mathbf{j}^{o_0} \mathbf{i}^{n_0}.
\end{equation*}
We can then put this into the normal ordering of  Eq.~(\ref{eq:pres}) using the relations in Eq.~(\ref{eq:genrel}) to find:
\begin{equation}
    g^{-1}= (-1)^{m_1} \mathbf{i}^{n_1} \mathbf{j}^{o_1} \mathbf{l}^{p_1 + 2q_1}.
\end{equation}
where the relation between the $|g\rangle$  and $|g^{-1}\rangle$ indices are :
\begin{equation}
\label{eq:inversionrules}
    \begin{split}
        m_1 &= m_0 + n_0 + o_0 + n_0 \times o_0\\
        n_1 &= n_0(1 - q_0) + o_0(p_0 + q_0)\\
        o_1 &= o_0(1 - p_0) + n_0(p_0 + q_0)\\
        p_1 &= q_0\\
        q_1 &= p_0.\\
    \end{split}
\end{equation}
A qubit circuit implementation of $\mathfrak U_{-1}$ is shown in Fig. \ref{fig:inversiongate}. We can map Eq. (\ref{eq:inversionrules}) onto a quantum circuit using modular arithmetic, finding that transforming $m_0$ to $m_1$ uses two CNOTs and a Toffoli gate. A circuit with a CSWAP and two Toffolis is required for $n_1$ and $o_1$, while $p_1$ and $q_1$ need one SWAP.

\begin{figure*}
 \includegraphics[width=0.9\linewidth]{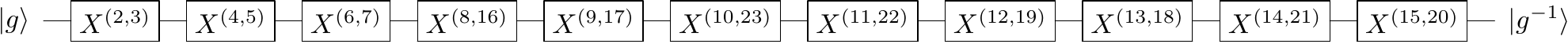}
    \caption{An \qit implementation of $\mathfrak U_{-1}$ using the $X^{(a,b)}$ gate.}
    \label{fig:quditinversion}
\end{figure*}

The \qit circuit of $\mathfrak U_{-1}$ is simpler, needing only 11 $X^{(a,b)}$ gates\footnote{1 $X^{(a,b)}$ of the 12 is unnecessary since $\ket{0^{-1}}=\ket{0}$ and $\ket{1^{-1}}=\ket{1}$}, as seen in Fig.~\ref{fig:quditinversion}. 24 $\mathcal{S}(\vec{\theta})$ and 25 $\mathcal{D}(\alpha)$ are sufficient to approximate $\mathfrak{U}_{-1}$ to sub-percent infidelity. Inspecting this circuit, the largest separation between inverses is $\ket{10}$ and $\ket{23}$.  Using only $\mathcal{S}(\vec{\theta})$ and $\mathcal{D}(\alpha)$, this could prove noisy in terms of the necessary $\vec{\theta}$ and $\alpha$, and thus large-separation photon blockade gates are desirable. 

\section{\label{sec:multiplication}Multiplication Gate}
\begin{figure}[!ht]
 \includegraphics[width=0.85\linewidth]{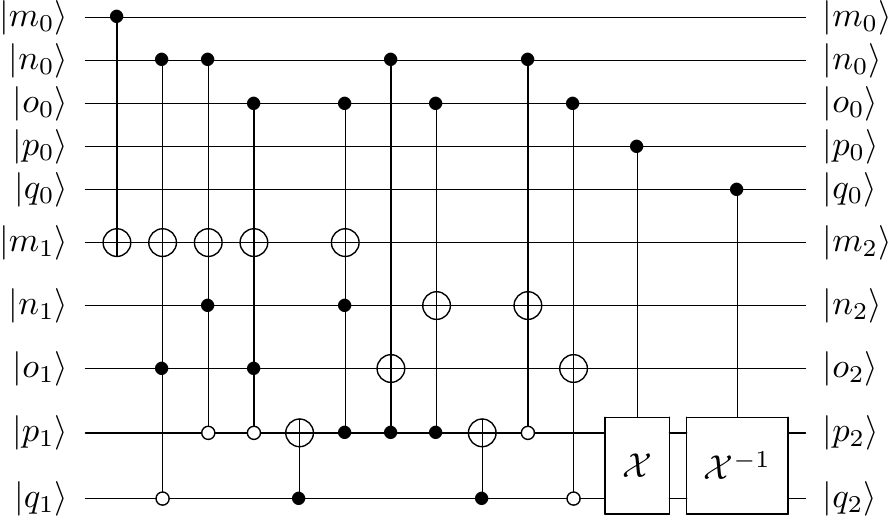}
    \caption{A qubit implementation of $\mathfrak U_{\times}$. Following convention, filled (open) circles correspond to control on $\ket{1}$ ($\ket{0}$).  The C$\chi$ gate is defined in Fig.~\ref{fig:dcpgate} and C$\chi^{-1}$ is its inverse.}
    \label{fig:multgate}
    \end{figure}

The method to construct the $\mathfrak U_{\times}$ for qubits is similar to that for $\mathfrak U_{-1}$. Given two $\btt-$registers storing $g$ and $h$:
\begin{align}
    g &= (-1)^{m_0} \mathbf{i}^{n_0} \mathbf{j}^{o_0} \mathbf{l}^{q_0 + 2p_0},\, h= (-1)^{m_1} \mathbf{i}^{n_1} \mathbf{j}^{o_1} \mathbf{l}^{q_1 + 2p_1},
\end{align}
we want $gh=g\times h$ and permuting $|h\rangle$ to $|gh\rangle$.  Defining $gh=(-1)^{m_2} i^{n_2} j^{o_2} l^{2p_2 + q_2}$, we can derive via Eq.~(\ref{eq:genrel}):
\begin{align}
        m_2 &= o_1n_0(1-p_1) + (n_1n_0 + o_1o_0)(1-q_1)\notag\\
        &\phantom{xxx}+n_1o_0(p_1+q_1)\notag\\
        n_2 &= n_1 + n_0(1-q_1) + o_0(p_1+q_1)\notag\\
        o_2 &= o_1+ o_0(1-p_1) + n_0(p_1 + q_1)\notag\\
        p_2 &= p_0(1-q_1)(1-p_1)\notag\\
        q_2 &= q_0(1-q_1)(1-p_1).
\end{align}
These expressions map into the qubit circuit of Fig.~\ref{fig:multgate}.

\begin{figure*}
  \includegraphics[width=0.8\linewidth]{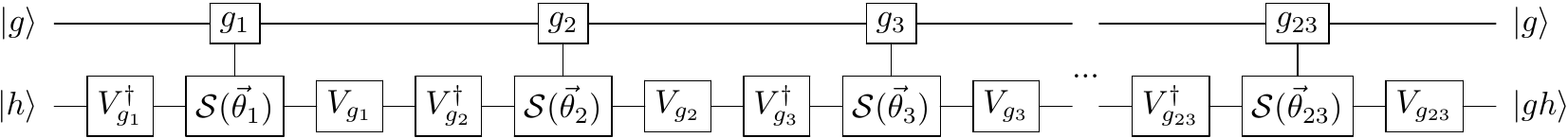}
    \caption{A \qit implementation of $\mathfrak U_{\times}$. The subscript $g_i$ indicates the the i$^{th}$ element of the group. }
    \label{fig:multiplicationqudit}
\end{figure*}

$\mathcal{U}_{\times}$ on \qits is a permutation  of $\ket{h}$ controlled by $\ket{g}$ realized as
\begin{equation*}
    \mathcal{U}_{\times} = \sum_{g \in G}\sum_{h \in G}|g\rangle \langle g|\otimes|h\rangle\langle g\times h| = \sum_{g\in G} |g\rangle\langle g|\otimes \hat{P}_{g},
\end{equation*}
where $\hat{P}_g$ is a permutation matrix that depends upon $g$. This unitary matrix can be diagonalized by one-qudit gates $V_g$. A \qit circuit for $\mathfrak U_{\times}$ is shown in Fig.~\ref{fig:multiplicationqudit}.

The structure of $V_g$ depends on the order $m$ of the element $g$ (Tab. \ref{tab:charbt}) being multiplied onto the element $h$. For a given operator $g$ the elements $h \in G$ will break down into $24/m$ sets of $m$ elements. The elements in each set are determined solely by the operator $g$ itself. The elements in each set can be generated by taking $h$ and left multiplying it by $g$ until the element $h$ is reached again. These sets of elements will provide a presentation of a $\mathbb{Z}_m$ group. In this way $V_g$ can be rendered into a set of $24/m$ blocks of size $m$. These blocked sections with at most an $SU(6)$ rotation in the given subspace. The $SU(N)$ Euler angle decompositions are provided in Appendix \ref{app:appendix} along with the group cycles generated by each $g$. Tab.~\ref{tab:givenscylcetab} provides the total number of $R_{p}^{(a,b)}(\theta)$ for each $QFT_{\mathbb{Z}_m}$ on the $m$-level subspace as well as $V_g$.

\begin{figure}[ht]
    \centering
    \includegraphics[width=0.49\linewidth]{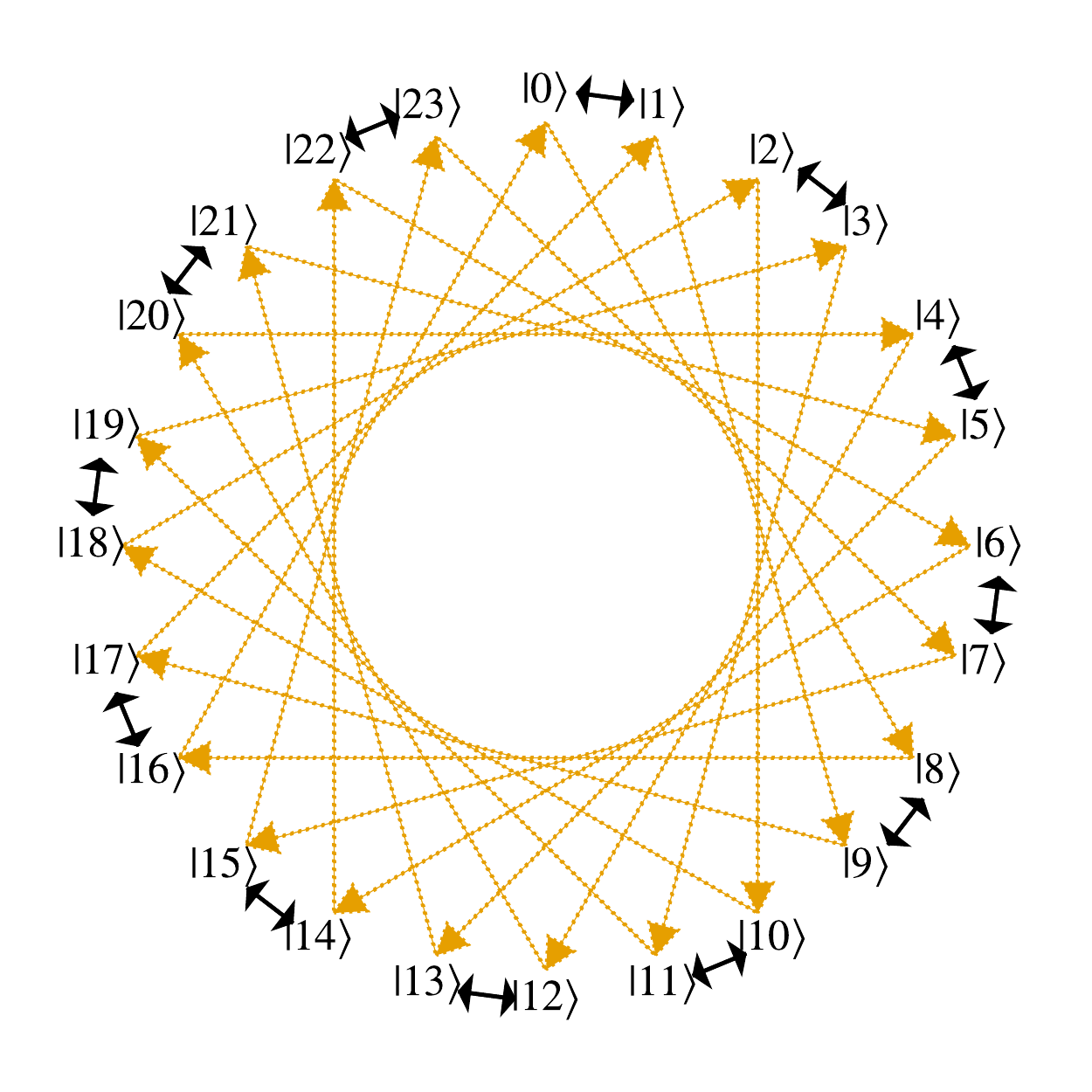}
    \includegraphics[width=0.49\linewidth]{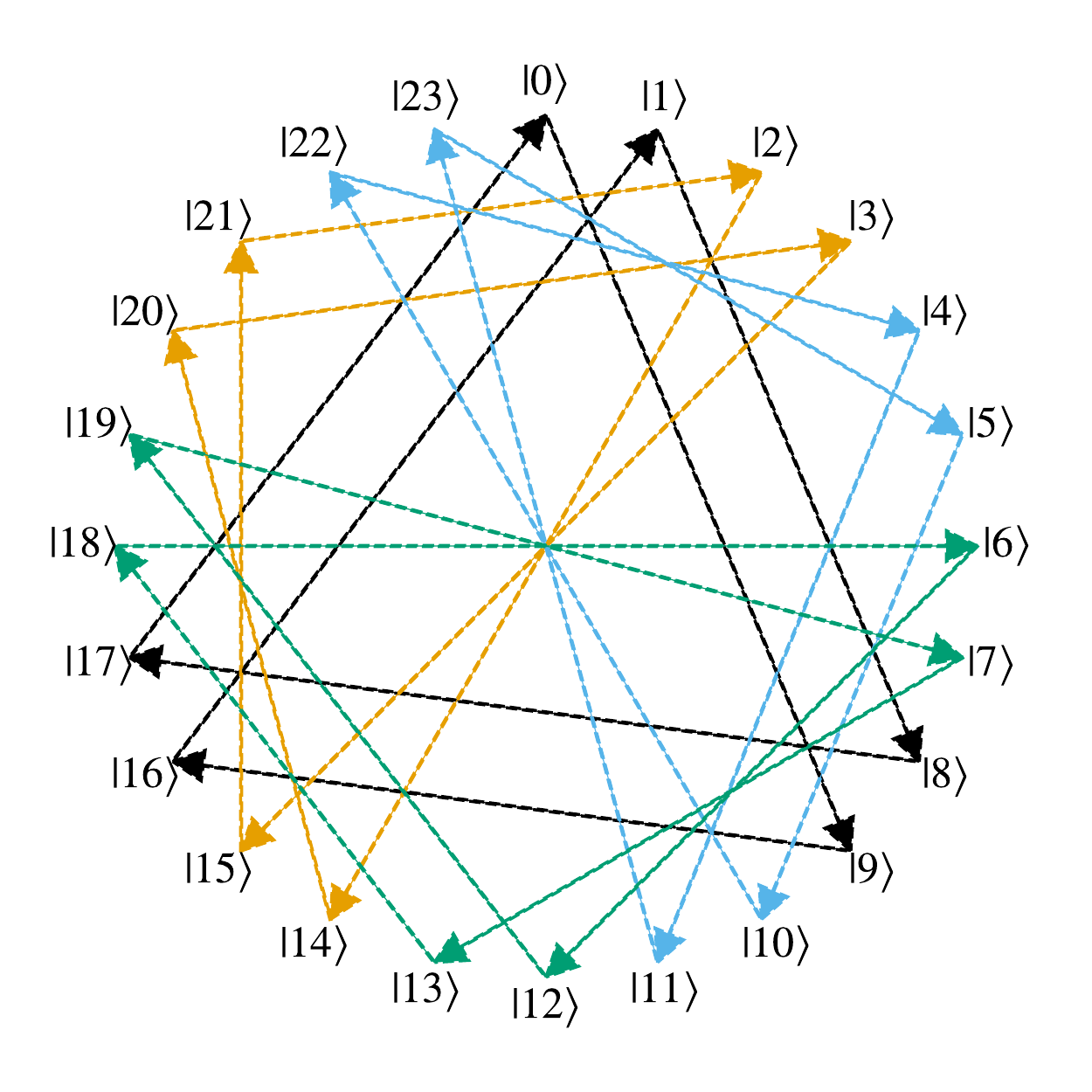}
    \caption{Left: Pictorial representation of the cycles for $\ket{g}$ being: $\ket{1}$ with $m=2$, (black arrows); $\ket{8}$ with $m=3$, (orange dotted arrows). Right: representation of the cycle for $\ket{9}$ with $m=6$, different cycles are shown in different colors.}
    \label{fig:multiplication_cycles}
\end{figure}

A directed graph for the group sets for $g=-1,~\mathbf{l},$ and $-\mathbf{l}$ ($\ket{1}$, $\ket{8}$, $\ket{9}$ respectively) are shown in Fig. \ref{fig:multiplication_cycles}. In this figure we show how multiplication of a group element $h$ on the left by a group element $g$ will cycle through a subset of the group elements in a directed graph. For example multiplication by $-1$ ($\ket{1}$) flip flops elements $|2a\rangle$ and $|2a+1\rangle$. In this way the neighboring states will have a $\mathbb{Z}_2$ Fourier transform applied on each pair. Multiplication by the element $\mathbf{l}$ ($\ket{8}$) will cycle the states $|a\rangle,~\ket{a + 8},$ and $\ket{a+16}$ for $a \leq 7$. The cycles shown for multiplication by  $-\mathbb{l}$ ($\ket{9}$), are more complicated to write in closed form but are shown in the right hand side of Fig. \ref{fig:multiplication_cycles}.
\begin{table}[ht]
    \caption{$R_{p}^{(a,b)}(\theta)$ required for $QFT_{\mathbb{Z}_m}$ and $V_g$ for each order. $N_{\ket{g}}$ denote the number of elements with that cycle.}
    \label{tab:givenscylcetab}
    \centering
    \begin{tabular}{cc|cc}
    \hline\hline
         cycle & $N_{\ket{g}}$ & $R_{p}^{(a,b)}(\theta)$ in $QFT_{\mathbb{Z}_m}$ & $R_{p}^{(a,b)}(\theta)$ in $V_g$\\
         \hline
         1 & 1 & 0 & 0\\
         2 & 1 & 3 & 36 \\
         3 & 8 & 8 & 64 \\
         4 & 6 & 15 & 90\\
         6 & 8 & 35 & 140\\
         \hline
    \end{tabular}
\end{table}

In this way $V^{\dagger}_g \hat{P}_g V_g$ will be a diagonal matrix whose nonzero elements are phases corresponding to the eigenvalues of $\hat{P}_g$. As we iterate through $g_i$, neighboring $V_{g_{i}} V^{\dagger}_{g_{i+1}}$ can be combine into a single qudit operation. If we use Tab.~\ref{tab:givenscylcetab} as a starting point and recognize that for the order $3$, $4$, and $6$ that $V_g$'s appear in pairs of states, almost half of the $R_{p}^{(a,b)}(\theta)$ are eliminated, leaving 2,244 to implement all $V_g$'s. In terms of native gates, $\mathfrak{U}_{\times}$ needs 575 SNAP and 575 $\mathcal{D}(\alpha)$ gates in addition to the 23 cSNAP gates for $\hat{P}_g$. This cost could be reduced by pulse engineering, an active research area in bosonic quantum computers~\cite{PhysRevLett.89.188301,opcon2106,Ozguler:2022itt,BarisOzguler:2022rwa}. 
    
\section{\label{sec:trace}Trace gate}
\begin{figure*}
 \includegraphics[width=\linewidth]{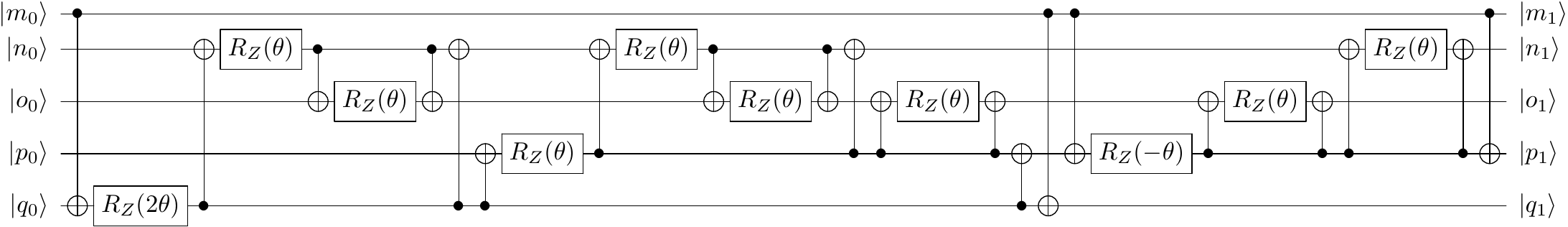}
    \caption{A qubit implementation of $\mathfrak U_{\rm Tr}$.}
    \label{fig:binarytetrahedralgatetrace}
\end{figure*}

For simulating gauge theories, $\mathfrak{U}_{\rm Tr}$ is only needed for the fundamental representation, $\rho_4$.  The character table (Tab.~\ref{tab:charbt}) provides us with the $\re\tr(g_i)$ necessary. $\mathfrak{U}_{\rm Tr}$ can be obtained by defining a Hamiltonian, and then exponentiating it. For our qubit-register, $H_{tr}$ for $\rho_4$ is
\begin{equation}
\begin{split}
    H_{\rm Tr} = & Z_{m_0} \bigg(Z_{q_0}\big[2 + \big(1 + Z_{o_0}\big)\big(Z_{n_0} + Z_{p_0}(1 + Z_{n_0})\big]\\
    &+ Z_{p_0}\big[Z_{o_0} + Z_{n_0} - 1\big]\bigg)
    \end{split}
\end{equation}
were $Z_h$ acts on the $|h\rangle$ qubit. From this, we can decomposing $e^{i\theta H_{tr}}$ into the linear combinations of $R_Z(\theta)$ gates. The qubit-based circuit for $\mathfrak U_{\rm Tr}$ is shown in Fig.~\ref{fig:binarytetrahedralgatetrace}. 

\begin{figure*}
 \includegraphics[width=0.9\linewidth]{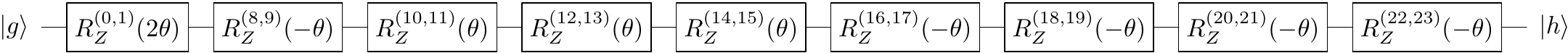}
    \caption{A \qit implementation of $\mathfrak U_{Tr}$ using two level $R_Z^{(a,b)}(\theta)$ gates.}
    \label{fig:quditrz}
\end{figure*}

Implementing $\mathfrak U_{\rm Tr}$ on a \qit requires 9 $R^{(a,b)}_Z(\theta)$ gates corresponding to the 9 ($g_i$,$-g_i$) pairs with $\re\tr(g_i)\neq0$; this gate is shown in Fig.~\ref{fig:quditrz}. Together, these gates can be mapped to a single SNAP gate.  

\section{\label{sec:fourier}Fourier Transform}
\begin{figure*}
    \centering
    \includegraphics[width=0.82\linewidth]{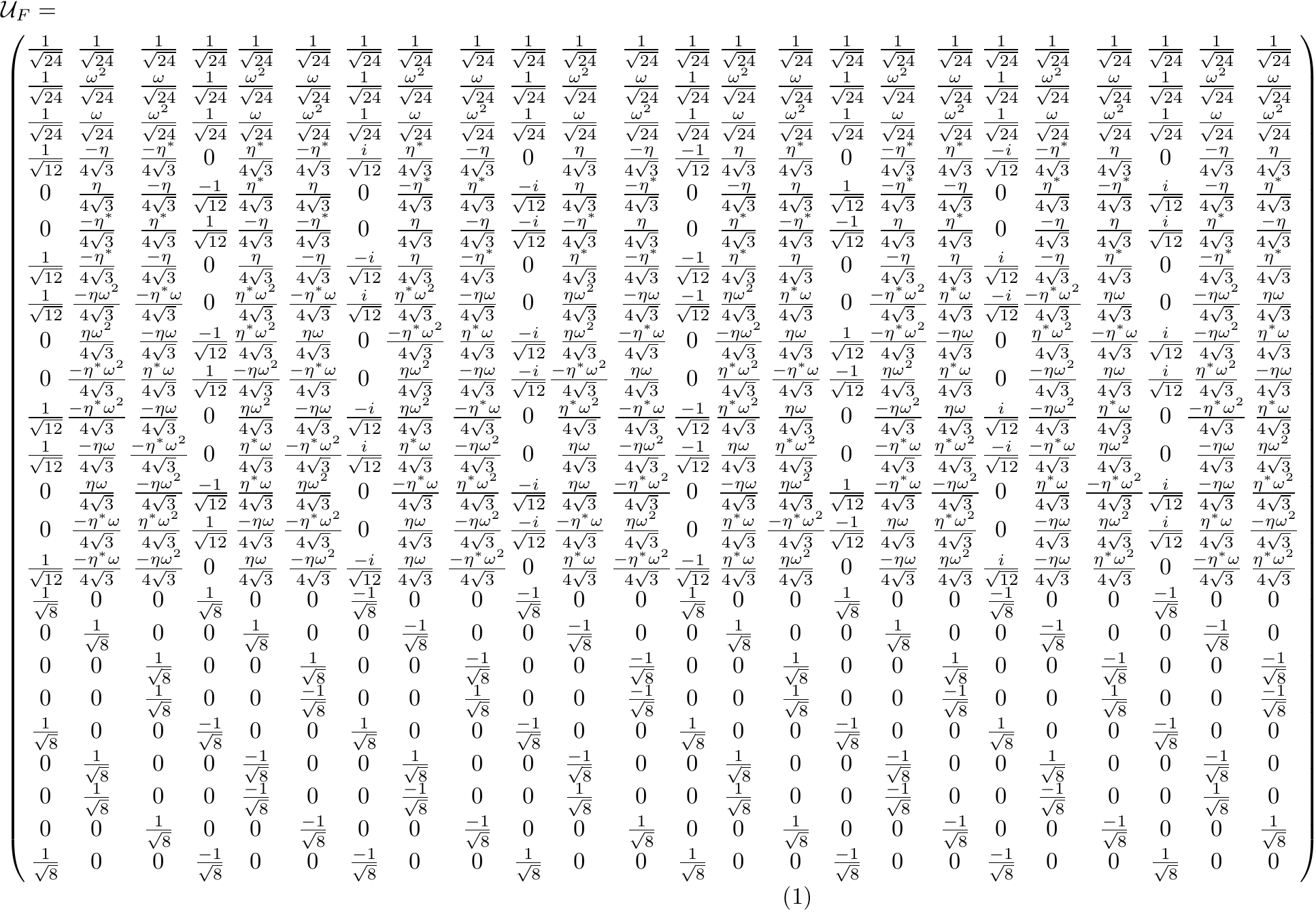}
    \caption{Matrix representation of $\mathfrak{U}_F$  where $\eta=1+i$ and $\omega=e^{\frac{2\pi i}{3}}$.}
    \label{fig:qft_mat}
\end{figure*}

The standard $n$-qubit quantum Fourier transform (QFT)~\cite{nielsen_chuang_2010} corresponds to the quantum version of the fast Fourier transform of $\mathbb{Z}_{2^n}$. 
Quantum Fouirer transforms over several nonabelian groups exist in the literature~\cite{hoyer1997efficient,beals1997quantum,puschel1999fast,moore2006generic,Alam:2021uuq}. Alas, for all the crystal-like subgroups of interest to high energy physics efficient QFT circuits are currently unknown~\cite{childs2010quantum}. For the general case, there isn't a clear algorithmic way to construct the QFT.  Therefore, we instead construct a suboptimal $\mathfrak U_{F}$ from Eq.~(\ref{eqn:Fourier-group}) using the irreps of Sec.~\ref{sec:group} to obtain the matrix in Fig.~\ref{fig:qft_mat}.

Since $\btt$ has 24 elements, on a qubit device $\mathfrak U_{F}$ must be embedded into a larger $2^d\times2^d$ matrix. With this matrix, a transpiler can be used to derive a circuit. Using the \textsc{Qiskit} transpiler, $\mathfrak U_{F}$ requires 1025 CNOTs, 2139 $R_Z(\theta)$, and 1109 $R_Y(\theta)$; the Fourier gate is the most expensive qubit primitive.  As will be discussed in Sec.~\ref{sec:resources}, $\mathfrak U_F$ dominates the total simulation costs and future work should be devoted to finding a $\btt$ QFT.

 A \qit implementation follows the same generalized Euler angle decomposition as the $V_g$ used for $\mathfrak{U}_{\times}$~\cite{2005quant.ph.11019S}.  This gate can be implemented with 24 SNAP and 25 displacement gates to subpercent infidelity.

\section{Experimental Results}
\label{sec:results}
\begin{figure}
    \includegraphics[width=0.65\linewidth]{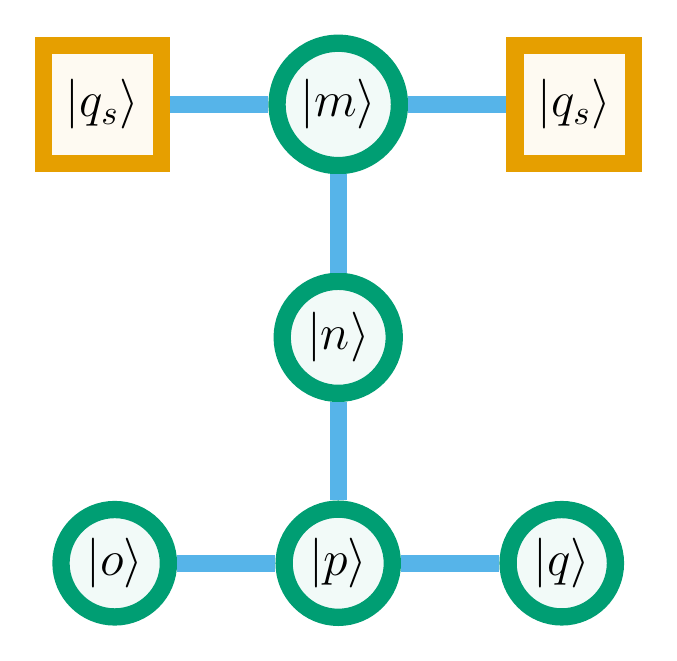}
    \caption{An example mapping of $\ket{g}=|qponm\rangle$ onto the 7 transmon qubit \texttt{ibm\_nairobi}}.
    \label{fig:nairobi}
\end{figure}

In this section, we discuss experimental results from running $\mathfrak U_{-1}$ and $\mathfrak U_{\rm Tr}$ on the \texttt{ibm\_nairobi} 7 transmon qubit device (see Fig.~\ref{fig:nairobi}). Transpiling Fig.~\ref{fig:inversiongate} and Fig.~\ref{fig:binarytetrahedralgatetrace} onto \texttt{ibm\_nairobi}, topology constraints require introducing additional SWAP gates (see Fig.~\ref{fig:nairobi}). The cost of $\mathfrak U_{\rm Tr}$ increases from 22 CNOTs to 39, and for $\mathfrak{U}_{-1}$ the number of CNOTs goes from 31 to 49. The high qubit cost of $\mathfrak U_{\times}$ and high gate costs of $\mathfrak U_{F}$ suggest they are unlikely to have reasonable fidelities and they are left for the future. 

We define the process fidelity $\mathcal{F}$ of $\mathfrak U$ on a state $\ket{\psi}$ as 
\begin{equation}
 \mathcal{F}^{|\Psi\rangle}_{\mathfrak{U}}=|\langle 0 |\Psi^{\dag}\mathfrak U ^{\dag}\mathfrak U \Psi|0\rangle|^2
\end{equation}

Without noise, the state preparation $\Psi$ and $\mathfrak{U}$ are exactly cancelled by their complex conjugations, thus the measured result should always be $\ket{0}^{\otimes5}$.
Determining the fidelity requires testing all the possible states $|\Psi\rangle$, a prohibitively expensive task~\cite{Chuang:1996hw}. Therefore we consider a subset of states given by the 24 group element states $|g\rangle$ which can be obtained by applying $X$ gates to the appropriate qubits. For a general state, $\Psi^{\dag}\mathfrak U ^{\dag}$ could require as many CNOTs as $\mathfrak U \Psi$.  In this case, the total circuit cost is doubled and fidelities are reduced.  In contrast, for $|g\rangle$, the results of applying either of our gates is another $|g\rangle$, which can be returned to $|0\rangle^{\otimes 5}$ using only $X$ gates.  With this, we compute $\mathcal{F}$ for each $\ket{g}$ for both $\mathfrak U_{\rm Tr}$ and $\mathfrak U_{-1}$ without doubling the CNOT count.  With these results we calculate a mean value $\bar{\mathcal{F}}_{\mathfrak U}$.

\begin{figure}
    \centering
    \includegraphics[width=\linewidth]{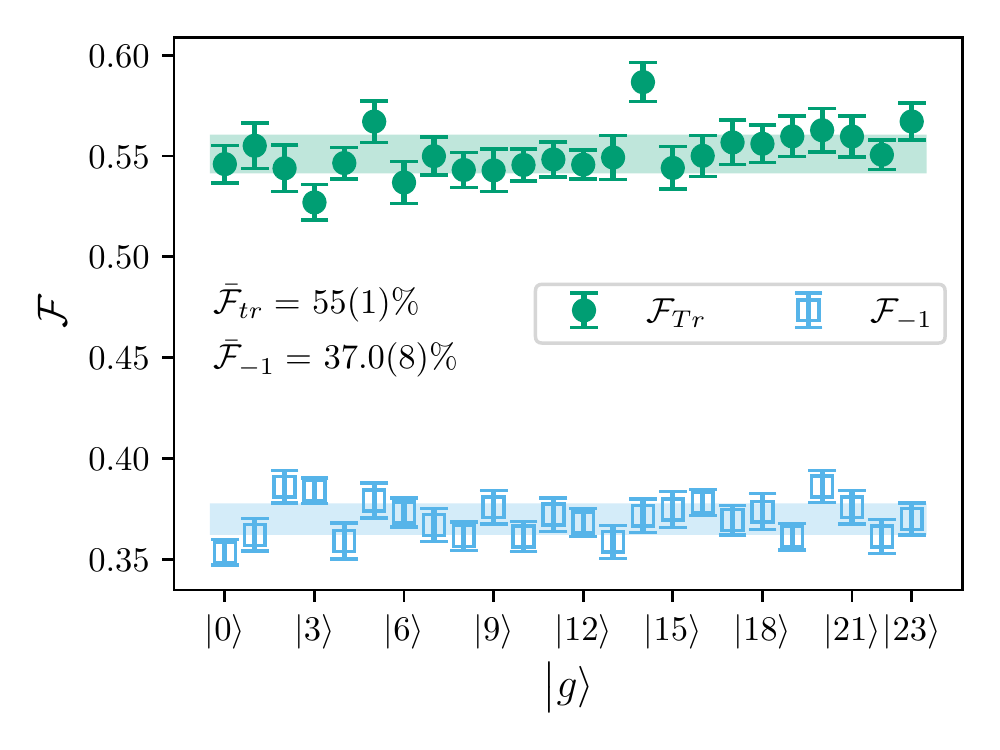}
    \caption{Process Fidelities, $\mathcal{F}$, for the trace and inverse gate on IBM's computer. The angle used was $\theta=0.7$. The averages are shown as a shaded band.}
    \label{fig:tracefidelity}
\end{figure}

The dominant coherent CNOT error can be mitigated through Pauli twirling~\cite{Erhard_2019,li2017efficient, 2018efficienttwirling, 2013efficienttwirl, 2016efficienttwirling}.  This method converts coherent errors into random Pauli channel errors and has found success in lattice applications~\cite{Yeter-Aydeniz:2022vuy,Carena:2022kpg}. The circuits are modified by wrapping each CNOT with a set of Pauli gates $\{\mathbb{1},X,Y,Z\}$ sampled from sets which are logical equivalent to CNOT. We ran 20 unique circuits for each gate-state pair following prior results finding $\mathcal{O}(10)$ circuits to suffice~\cite{Erhard_2019}.  2000 shots were taken for the $\mathfrak{U}_{\rm Tr}$ circuits, while 500 were gotten for $\mathfrak{U}_{-1}$. The process fidelity for each $\ket{g}$ are shown in Fig.~\ref{fig:tracefidelity}. Averaging we find that the $\bar{\mathcal{F}}^{|g\rangle}_{\rm Tr}=55(1)\%$, while the higher CNOT count of $\mathfrak{U}_{-1}$ leads to a lower fidelity of $\bar{\mathcal{F}}^{|g\rangle}_{-1}=37.0(8)\%$.

While $\ket{g}$ are easy to implement, they are less likely to be encountered during a simulations, since states must be gauge invariant.  In the case of $\mathfrak{U}_{-1}$, it is possible to test a gauge invariant state $|GI\rangle=|G|^{-1/2}\sum_{g}|g\rangle$ because both $|GI\rangle$ and  $\Psi^{\dag}\mathfrak{U}_{-1}^{\dag}$ can be implemented with only 1 additional CNOT. For this state, we found a reduced fidelity $\mathcal{F}^{|GI\rangle}_{-1}=14.8(12)\%$.

\section{\label{sec:resources}Resource Estimates}

Clearly time evolution for large lattices of $\btt$ is beyond the NISQ era and we should consider fault-tolerant quantum computers. The Eastin-Knill theorem restricts quantum error correcting codes by preventing universal sets of gates from being implemented transversally~\cite{Eastin_2009}.
Transversality refers to the property that gates operating on logical qubits correspond to tensor products on the physical qubits. For many error correcting codes such as Calderbank-Shor-Steane (CSS) codes the Clifford gates are transversal \cite{Chuang:1996hw,1996PhRvA..54.1098C,PhysRevLett.77.793,1996RSPSA.452.2551S,PhysRevA.54.4741} while the T gate is not. Therefore, T gates are an important metric in fault-tolerant algorithm analysis because they require entanglement between physical qubits~\cite{Chuang:1996hw,1997RuMaS..52.1191K}. 

\begin{table}[]
    \centering
    \caption{Number of physical T gates and clean ancilla required to implement logical gates for (top) basic gates taken from~\cite{Chuang:1996hw} (bottom) primitive gates for $\btt$.}
    \label{tab:tgatecost}
    \begin{tabular}{ccc}
    \hline\hline
         Gate & T gates & Clean ancilla\\
         \hline
         C$^2$NOT & 7 & 0\\
         C$^3$NOT & 28 & 1\\
         CSWAP & 7 & 0\\
         $R_z$ & 1.15 $\log_2(1/\epsilon)$ & 0\\
         \hline
         $\mathcal{U}_{-1}$ & 28 & 0\\
         $\mathcal{U}_{\times}$ & 154 & 1\\
         $\mathcal{U}_{Tr}$ & 12.65 $\log_2(1/\epsilon)$ & 0\\
         $\mathcal{U}_{FT}$ & 1150 $\log_2(1/\epsilon)$ & 0\\\hline\hline
    \end{tabular}
\end{table}

The Toffoli gate is known to require 7 T gates~\cite{Chuang:1996hw} and the C$^n$NOT gates can be constructed exactly using a ladder of Toffoli gates and clean ancilla qubits\footnote{A \emph{clean} ancilla is a qubit initialized to $\ket{0}$. \emph{Dirty} ancilla indicate ones in an unknown initial state.} which can be reused later~\cite{Chuang:1996hw,PhysRevA.52.3457}. Using this ladder method a $C^n$NOT gate can be implemented using $4(n-1)$ Toffoli gates and $n-1$ clean ancilla qubits. Methods exist using dirty ancilla at the cost of more T gates~\cite{2019arXiv190401671B, PhysRevA.52.3457}.  We arrive at the cost for the $R_z$ gates via \cite{PhysRevLett.114.080502} where these gates can be approximated to precision $\epsilon$ with at worst $1.15 \log(1/\epsilon)$ T gates using the repeat-until-success method.  Using these, we can construct fault-tolerant gate cost estimates for $\btt$ (See Tab.~\ref{tab:tgatecost}).

\begin{table}[ht]
   \caption{Number of primitive gates per link per $\delta t$ neglecting boundary effects as a function of dimension $d$ for $H_{I}$.}
    \label{tab:primcost}
    \begin{tabular}{c|c}
    \hline\hline
    Gate &$N[H_{I}]$\\
    \hline
    $\mathfrak U_F$ &4\\
    $\mathfrak U_{\rm Tr}$&$\frac{3}{2}(d-1)$\\
    $\mathfrak U_{-1}$& $2+11(d-1)$\\
    $\mathfrak U_{\times}$&$4+26(d-1)$\\
    \hline\hline
    \end{tabular}
\end{table}
Primitive gate costs for implementing the improved Hamiltonian, $H_I$, per link per Trotter step $\delta t$ are shown in Tab.~\ref{tab:primcost}. Using these costs we find that a $d$ spatial lattice simulation of $H_{I}$ for time $t=N_t\delta t$ would require
\begin{equation}
    N_{T}=\left[4312d-3640+(4581.03+18.975d\log_2\frac{1}{\epsilon})
    \right] d L^d N_t
\end{equation}
Following \cite{Cohen:2021imf,Kan:2021xfc}, we consider a fiducial simulation of the shear viscosity $\eta$ on a $d=3$ lattice of $L=10$ with $N_t=50$, $\epsilon=10^{-8}$ and the cost of state preparation neglected. For an $SU(2)$ simulation including fermions, Kan and Nam estimated $3\times10^{34}$ T gates, while neglecting fermions allows for a more modest $3\times10^{19}$. Here, we neglect fermions and using $\mathbb{BT}$ to approximate $SU(2)$ requires $2.0\times10^{10}$ T gates for $H_I$. So using $\btt$ reduces the gate costs by 9 orders of magnitude. The T gate density is 1 T gate per $\btt-$register per clock cycle and is independent of primitive, although a QFT might increase this.  The large reduction in  T gates compared to \cite{Kan:2021xfc} comes by avoiding quantum fixed-point arithmetic.  For us, $\mathfrak{U}_{F}$ dominates the simulations -- $44\%$ of the total cost.

Compared to qubits, the field of quantum error correction for qudits is less developed~\cite{959288,1715533,luo2017non,galindo2019entanglement,9207971,PhysRevA.73.032325,imany2019high,https://doi.org/10.48550/arxiv.1906.11137,Galindo_2020,PhysRevA.103.042420,https://doi.org/10.48550/arxiv.2008.00713,chizzini2022quantum,PhysRevA.97.052302,https://doi.org/10.48550/arxiv.2202.09235,Vandermolen:2021ryw}.  Much, but not all, of the work has focused on qutrits and relies upon specific hardware and native gates.  While this field will develop rapidly in the coming years, we will restrict ourselves to \qit resources estimates based on a device with native cSNAP, SNAP, and displacement gates.  The costs for each $\btt$ gate are shown in Tab.~\ref{tab:qitcost}. In contrast to the qubits, for \qit simulations the most costly gate is $\mathfrak{U}_{\times}$ with all other gates contributing negligible amounts. Thus, determination of the QFT is less important for \qit devices.  Summing the gates, we find the fiducial calculation of the viscosity with 3$\times10^3$ \qits would require $1.9\times10^8$ cSNAP and $4.9\times 10^9$ SNAP and displacement gates.

\begin{table}[]
    \centering
    \caption{c$\mathcal{S}(\vec{\theta})$, $\mathcal{S}(\vec{\theta})$, and $\mathcal{D}(\alpha)$ gates required for $\btt$ (top) primitive gates (bottom) $H_{I}$ simulations per link per $\delta t$.}
    \label{tab:qitcost}
    \begin{tabular}{cccc}
    \hline\hline
         Gate & c$\mathcal{S}(\vec{\theta})$ & $\mathcal{S}(\vec{\theta})$ & $\mathcal{D}(\alpha)$\\
         \hline
         $\mathcal{U}_{-1}$ & 0 & 24 & 25\\
         $\mathcal{U}_{\times}$ & 23 & 575 & 575\\
         $\mathcal{U}_{Tr}$ & 0 & 1 & 0\\
         $\mathcal{U}_{FT}$ & 0 & 24 & 25\\
         \hline
         $e^{-iH_{I}\delta t}$ & $598d-506$ & $15215.5d-12771.5$ & $15225d-12775$\\
         \hline\hline
    \end{tabular}
\end{table}

\section{Conclusions}
\label{sec:conclusions}
In this paper, we constructed the necessary primitive quantum circuits for the simulation of $\btt$ -- the smallest crystal-like subgroup of $SU(2)$ -- gauge theories.  These circuits were constructed for both qubit and \qit architectures and quantum resource estimates were made for the simulation of pure $SU(2)$ shear viscosity.  Compared to previous fault-tolerant qubit estimates, we require $10^{9}$ fewer T gates by avoiding quantum fixed point arithmetic via the discrete group approximation. While these simulations are still far off, we performed quantum fidelity experiments for two of the gates. Experimentally, we found the fidelity of the inversion and trace operation to be $37.0(8)\%$ and $55(1)\%$ for classical bit string states on the \texttt{ibm\_nairobi} quantum processor.  

Qudit-based quantum computers, like the \qit device considered here, are known to require fewer gates, in particular entangling ones.  Here we have demonstrated an additional benefit that the construction of nonabelian group primitives are dramatically simplified compared to the qubit case by reducing the complex internal $G-$register logic required to preserve group structure.

Looking forward, primitive gates should be constructed for larger crystal-like subgroups of $SU(2)$ and to the subgroups of $SU(3)$ theories.  At the cost of more qubits and larger lattice errors, a larger $SU(2)$ subgroup should allow the possibility of using the Kogut-Susskind Hamiltonian. This would reduce gate costs by a factor of 2 on a qubit device and a factor of 4 on a qudit one since different primitives dominate the cost. Finally, in order to further reduce the qubit-based simulation gate costs for all discrete subgroup approximations, the formalism for deriving the quantum Fourier transform for each crystal-like subgroup would be of great interest.

\begin{acknowledgements}
The authors thank Do\u{g}a K\"urk\c{c}\"uo\u{g}lu and Sophie Croll for helpful comments.
EG is supported by the U.S. Department of Energy, Office of Science, National Quantum Information Science Research Centers, Superconducting Quantum Materials and Systems Center (SQMS) under contract number DE-AC02-07CH11359. HL and FL are supported by the Department of Energy through the Fermilab QuantiSED program in the area of ``Intersections of QIS and Theoretical Particle Physics". Fermilab is operated by Fermi Research Alliance, LLC under contract number DE-AC02-07CH11359 with the United States Department of Energy.  We acknowledge use of the IBM Q for this work. The views expressed are those of the authors and do not reflect the official policy or position of IBM or the IBM Q team.

\end{acknowledgements}
\bibliography{ref}

\appendix

\section{SU(N) Euler angle decompositions}
\label{app:appendix}
The following operators $\mathcal{U}_{2}^{(a,b)}$, $\mathcal{U}_{3}^{(a, b, c)}$, $\mathcal{U}_4^{(a, b, c, d)}$, and $\mathcal{U}_6^{(a, b, c, d, e, f)}$ correspond to specific $SU(N)$ rotations that implement the $QFT_{\mathbb{Z}_m}$ of Tab. \ref{tab:givenscylcetab}.
We use the following Euler angle decompositions where the superscripts indicate levels that are swapped between. The $SU(2)$ Euler angle decomposition we require is built from the well-known ZXZ rotation
\begin{equation}
    \mathcal{U}^{(a,b)}_2(\vec{\theta}_{II}) = R^{(a,b)}_Z(\theta_0) R^{(a,b)}_X(\theta_1)R^{(a,b)}_Z(\theta_2).
\end{equation}
For the two-state rotation we need, $\vec{\theta}_{II}=[\pi / 2,\pi / 2,\pi / 2]$. An example of the operator $V_g$ with cycle $m=2$ would be for $g=-\mathbb{1}$, which corresponds to $|1\rangle$, and is given by:
\begin{equation}
    \label{eq:vgm1}
    V_{1} = \prod_{a=0}^{11} \mathcal{U}_{2}^{(2a, 2a + 1)}(\vec{\theta}_{II}).
\end{equation}

The $SU(3)$ Euler angle decomposition requires two $\mathcal{U}^{(a,b)}_2(\vec{\theta})$ and two Givens rotations~\cite{2002JPhA...3510467T, 2006JMP....47d3510B, 2004JGP....52..263T, 2005quant.ph.11019S},
\begin{equation}
\begin{split}
    \mathcal{U}^{(a,b,c)}_3(\vec{\theta}_{III}) = & \mathcal{U}^{(a,b)}_2(\vec{\theta_0})
    R^{(b,c)}_X(\theta_1)\mathcal{U}^{(a,b)}_2(\vec{\theta_2})
     R_Z^{(b,c)}(\theta_3),
     \end{split}
\end{equation}
where the angles are fixed to: $\vec{\theta}_0=[7\pi/6,3\pi/2,\pi/2]$, $\theta_1=0.608175\pi$, $\vec{\theta}_4 = [0,-\pi/2,\pi/3]$, $\theta_3=7\pi/3$.
One element with $m=3$ is $g=\mathbf{l}$, corresponding to $\ket{8}$:
\begin{equation}
    V_{8} = \prod_{a=0}^{7} \mathcal{U}_3^{(a, a + 8, a + 16)}(\vec{\theta}_{III}).
\end{equation}
The Euler angle decomposition of an arbitrary $SU(4)$ is given in terms of three $\mathcal{U}^{(a,b)}_2(\vec{\theta})$ and six Givens rotations
\begin{equation}
\begin{split}
    \mathcal{U}^{(a,b,c,d)}_4(\vec{\theta}_{IV}) = &\, \mathcal{U}^{(a,b)}_2(\vec{\theta_0})
    R^{(b,c)}_X(\theta_1)R^{(a,b)}_Z(\theta_2) \\
     &R^{(c, d)}_Z(\theta_3)\mathcal{U}^{(a,b)}_2(\vec{\theta_4}) R^{(b,c)}_X(\theta_5)\\
     &\mathcal{U}^{(a,b)}_2(\vec{\theta_6}) R^{(b, c)}_Z(\theta_{7})R^{(c, d)}_Z(\theta_{8}),
     \end{split}
\end{equation}
where the angles required for $QFT_{\mathbb{Z}_4}$ are fixed to be: $\vec{\theta}_0=[2\pi,\pi/2,0]$, $\theta_1=1.392\pi$, $\theta_2 = 0.4511\pi$, $\theta_3 = 4 \pi / 3$, $\vec{\theta}_4 =[ 0.90126 \pi, 0.41956\pi, 1.852\pi]$, $\theta_5=0.60817\pi$, $\vec{\theta}_{6} = [\pi / 2,\pi / 4, -\pi / 4]$,
$\theta_{7} = -\pi / 2$,
$\theta_{8} = -3\pi/4$.
If we consider the example $g=\mathbf{i}$, ($\ket{2}$) then $V_{2}$ would be:
\begin{equation}
    V_{2} = \prod_{a=0}^{5} \mathcal{U}_4^{(4a, 4a+2, 4a+1, 4a+3)}(\vec{\theta}_{IV}).
\end{equation}

The final decomposition required for $\btt$ is $SU(6)$, which is given by two $\mathcal{U}^{(a,b,c)}_3(\vec{\theta})$ and three $\mathcal{U}^{(a,b)}_2(\vec{\theta_0})$:
\begin{align}
    \mathcal{U}_6^{(a,b,c,d,e,f)}=&\mathcal{U}_3^{(a,b,c)}(\vec{\theta}_{III})\mathcal{U}_3^{(d,e,f)}(\vec{\theta}_{III})\notag\\&\times\mathcal{U}_2^{(a,d)}(\vec{\theta}_{II})\mathcal{U}_2^{(b,e)}(\vec{\theta}_{II})\mathcal{U}_2^{(c,f)}(\vec{\theta}_{II}).
\end{align}
The group element $-\mathbf{l}$ corresponding to $\ket{9}$ has order $m=6$. The corresponding $V_9$ is made with four products of $\mathcal{U}_{6}^{(a, b, c, d, e, f)}$, although it lacks the obvious structure of the other examples shown thus far:
\begin{equation}
\begin{split} 
    V_{9} = & \mathcal{U}_{6}^{(0, 9, 16, 1, 8, 17)} \mathcal{U}_{6}^{(2, 14, 20, 3, 15, 21)}\\
    &\mathcal{U}_{6}^{(4, 11, 23, 5, 10, 22)} \mathcal{U}_{6}^{(6, 12, 19, 7, 13, 18)}.
    \end{split}
\end{equation}
\end{document}